\documentclass[12pt,a4paper]{article}
\usepackage{graphicx}
\usepackage{times}
\newcommand\arcdeg{\mbox{$^\circ$}}%
\newcommand\arcmin{\mbox{$^\prime$}}%
\newcommand\arcsec{\mbox{$^{\prime\prime}$}}%

\newcommand\farcmin{\mbox{$.\mkern-4mu^\prime$}}%
\newcommand\farcsec{\mbox{$.\!\!^{\prime\prime}$}}%
\newcommand\micron{\mbox{$\mu$m}}%

\title{\bf Scientific potential of the Indo-Belgian 3.6-m DOT in the field of Galactic Astronomy
    \footnote{A potential footnote to the title may be added here}}

\author{Ram Sagar$^{1,2}$\thanks{ram\_sagar0@yahoo.co.in,ramsagar@iiap.res.in}, Brijesh Kumar$^2$, Annapurni Subramaniam$^1$\\
\vspace{0.5cm}\\
\normalsize $^1$ Indian Institute of Astrophysics, Bangalore, 560034\\ 
\normalsize $^2$ Aryabhatta Research Institute of Observational Sciences (ARIES), Manora Peak, Nainital, 263002 \\ }
\date{\mbox{}}
\begin{document}
\maketitle
\setcounter{page}{1001}
\pagestyle{plain}
    \makeatletter
    \renewcommand*{\pagenumbering}[1]{%
       \gdef\thepage{\csname @#1\endcsname\c@page}%
    }
    \makeatother
\pagenumbering{arabic}

%
%
\def\bull{\vrule height .9ex width .8ex depth -.1ex}
\makeatletter
\def\ps@plain{\let\@mkboth\gobbletwo
\def\@oddhead{}\def\@oddfoot{\hfil\scriptsize\bull\quad
"2nd Belgo-Indian Network for Astronomy \& astrophysics (BINA) workshop'', held in Brussels (Belgium), 9-12 October 2018 \quad\bull}%
\def\@evenhead{}\let\@evenfoot\@oddfoot}
\makeatother
%
%
\def\beginrefer{\section*{References}%
\begin{quotation}\mbox{}\par}
\def\refer#1\par{{\setlength{\parindent}{-\leftmargin}\indent#1\par}}
\def\endrefer{\end{quotation}}
%
%
{\noindent\small{\bf Abstract:} 
India and Belgium have jointly established two 4 meter class optical telescopes at Devasthal located in Nainital, India. 
After successful installation of the 3.6-m modern new technology Devasthal Optical 
Telescope (DOT) in 2015, it was technically activated by premiers of both countries from Brussels on March 30, 2016. 
Since then, the 3.6-m DOT has been used for both optical and near-Infrared (NIR) observations for a number of research 
proposals. The best angular resolution achieved is 0\farcsec4 indicating that the optics of the 3.6-m DOT 
is good and capable of providing images of the celestial bodies with sub-arc-second resolution. The observations 
provide proof that the care taken in the construction of the telescope dome building has paid a rich dividend as their 
thermal mass is so low that it has not degraded the natural atmospheric seeing at Devasthal measured about two decades 
ago during 1997 to 1999 using differential image motion monitor. 
  
A few preliminary scientific results obtained from recent observations are presented along with performance and global 
potential of the 3.6-m DOT in the field of galactic astronomy. The 3.6-m DOT is capable of providing 
internationally competitive science once high resolution spectrograph and other planned modern back-end instruments 
become operational. Geographical location of the observatory has global importance for the time domain and 
multi-wavelength astrophysical studies. 
}
\vspace{0.5cm}
\section{Introduction}
In order to have a long term scientific and technical collaboration between the scientists and engineers of Belgium and 
India, competent authorities of the two countries namely Belgian Federal Office for Science Policy (BELSPO) from 
Belgium and Department of Science \& Technology (DST) from India signed memorandum of understanding (MoU) on 
November 3, 2006 in New Delhi. The agreement covers a wide range of scientific fields 
including Physics and Astrophysics as one key area. Under this umbrella, the {\bf Belgo-Indian Network for Astronomy 
and Astrophysics (BINA)} has been created recently. It is aimed not only to foster the interaction between Indian 
and Belgian astronomers but also to develop jointly modern observing instruments in the field of Astronomy and 
Astrophysics. More details of the BINA collaboration are given by De cat et al. (2018) and Sagar (2018) along with 
scientific programmes and summary of the first BINA workshop.

An extended site survey for optical observations identified Devasthal (Longitude = 79.68\arcdeg\ E; Latitude = 29.36\arcdeg\ N; 
Altitude= 2424$\pm$4 meter) located in central Himalayan region of Kumaon, Uttarakhand, as a potential site for 
installing 4 meter class optical telescopes. Results of detailed characterization using modern equipment are published 
by Pant et al. (1999), Sagar et al. (2000 a, b) and Stalin et al. (2001). There are $\sim$ 210 spectroscopic nights in a 
year out of which about 80\% are of photometric quality. Micro-thermal measurements carried out during 1997 to 1998 
indicate that seeing near top of the Devasthal peak can be sub-arc-second for a good fraction of  observing time if 
primary mirror of the telescope is located at a height of about 13 meters or above from the ground.  

Sagar et al. (2019) have given a historical prospective the 3.6-m Devasthal optical telescope (DOT) project 
along with its sky performance. In brief, on 29th March 2007, the Advanced Mechanical Optical System (AMOS), Liege, 
Belgium was awarded the contract to design, manufacture, integration, testing, supply and installation of a 3.6 meter 
aperture size modern optical telescope at Devasthal. The BELSPO, Government of Belgium contributed 2 Million Euros 
towards the cost of the 3.6-m DOT project and in return secured 7\% of the telescope observing time for astronomical 
community of Belgium (De cat et al. 2018). The DST, Government of India (GoI) provided the remaining cost ($\sim$ 1250 
million Indian Rupees) of the telescope project through the Aryabhatta Research Institute of Observational Sciences (ARIES). 
Operation and maintenance of the 3.6-m DOT is the responsibility of the ARIES, Nainital through DST, GoI.

Details of the 3.6-m DOT opto-mechanical design are given by Flebus et al. (2008). The AMOS sub-contracted surface 
polishing of both primary (M1) and secondary (M2) mirrors to the Lytkarino Optical Glass Factory, Moscow, Russia who completed 
the task in 2010 and published as built optical quality of the mirrors (Semenov 2012). The AMOS in collaboration 
with ARIES, carried out factory assembly, integration and first light verification tests of the 3.6-m telescope in 
2012 and reported its workshop performance (Ninane et al. 2012). Technical details of mirror aluminum coating plant are 
given by Bheemireddy et al. (2016) while construction of modern telescope house and associated buildings, telescope 
assembly, installation and sky performance are described by Kumar et al. (2018) and Sagar et al. (2019). 

Figure 1 shows a long-distance view of the 1.3-m Devasthal Fast optical telescope (DFOT) (Sagar et al. 2011), 
3.6-m DOT and 4-m international liquid mirror telescope (ILMT) buildings located near Devasthal mountain peak. 
Present status of the 4-m ILMT has been documented by Surdej et al. (2018). It is in collaboration with 
researchers from various academic institutions of Belgium, Canada, India, Poland and Uzbekistan. As the ILMT has 
not yet seen its first light, only potential of the 3.6-m DOT for galactic astrophysics based on its sky 
performance is presented hereafter. After successful installation and performance verification, the 3.6-m DOT 
was technically activated jointly by both Indian and Belgian premiers on March 30, 2016. Since then, the telescope is in 
regular use for carrying out optical and near-infrared (NIR) observations of the celestial objects. Observing proposals 
are submitted online at the website: {\it http://www.aries.res.in/dopses/ } for both observing cycles named Cycle 1 
(February to May) and Cycle 2 (October to January). The policies and procedures followed in allotment of observing 
time for the 3.6-m DOT are given on the website {\it  http://www.aries.res.in/dot.} It also provides useful 
information regarding logistics, mode of observations and data rights etc. Based on the scientific merit of the 
submitted proposals, observing time is allotted by the Belgian and Indian time allocation committees for the 
proposers of their countries. Observing time allotted to proposers from ARIES and Belgium are presently 33\% and 7\% 
respectively while reaming 60\% observing time is allotted to proposers of other Indian Institutions.

\begin{figure*} 
\centering
\includegraphics[scale=0.145]{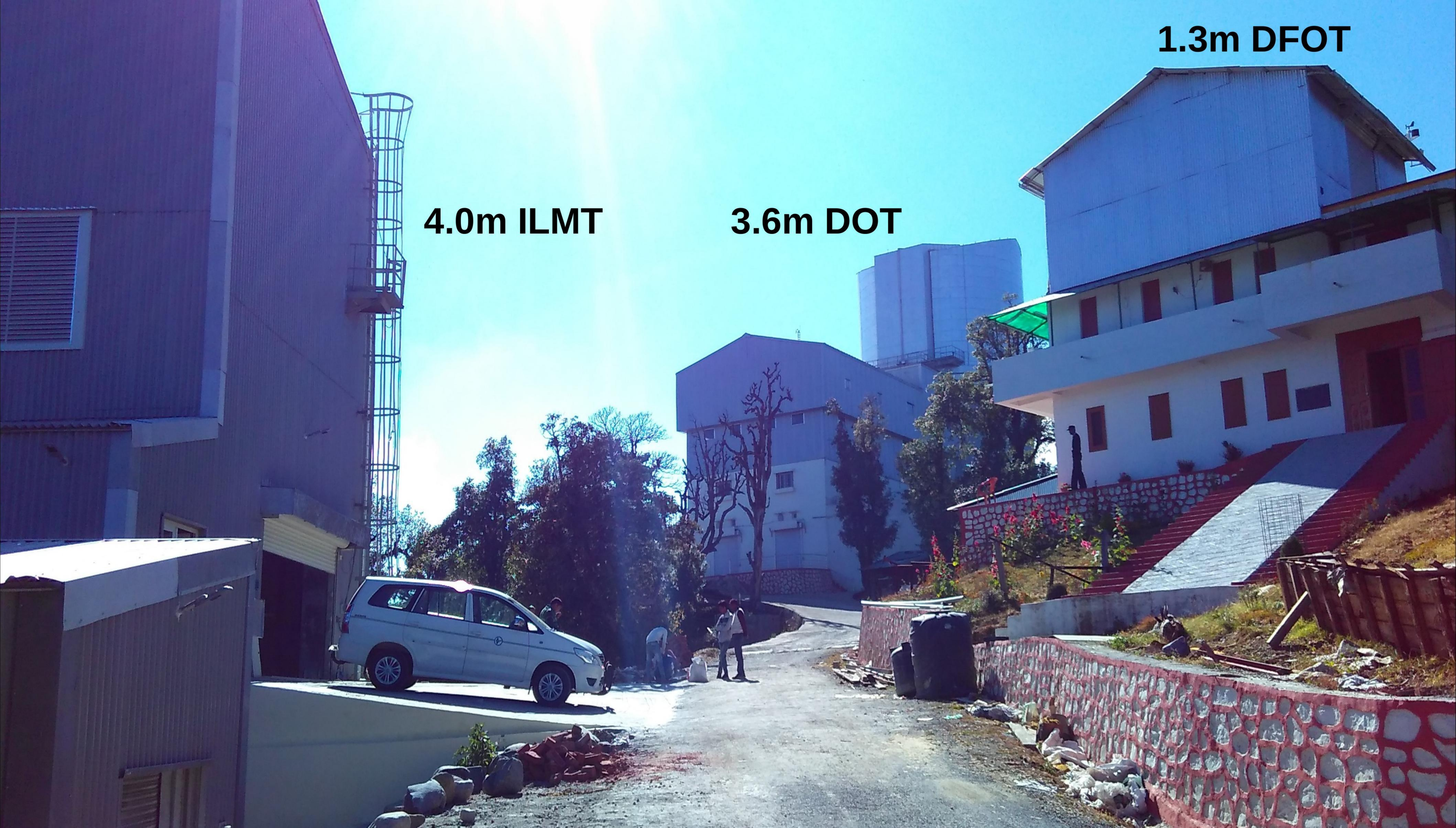}
\caption{Near top of the Devasthal mountain peak, locations of the 1.3-m DFOT, 3.6-m DOT and 4-m ILMT are shown. \label{fig_6}}
\end{figure*}
 
\section{Sky performance of the 3.6-m DOT and Back-end instruments}

\begin{figure}[h]
\centering
\includegraphics[width=16cm]{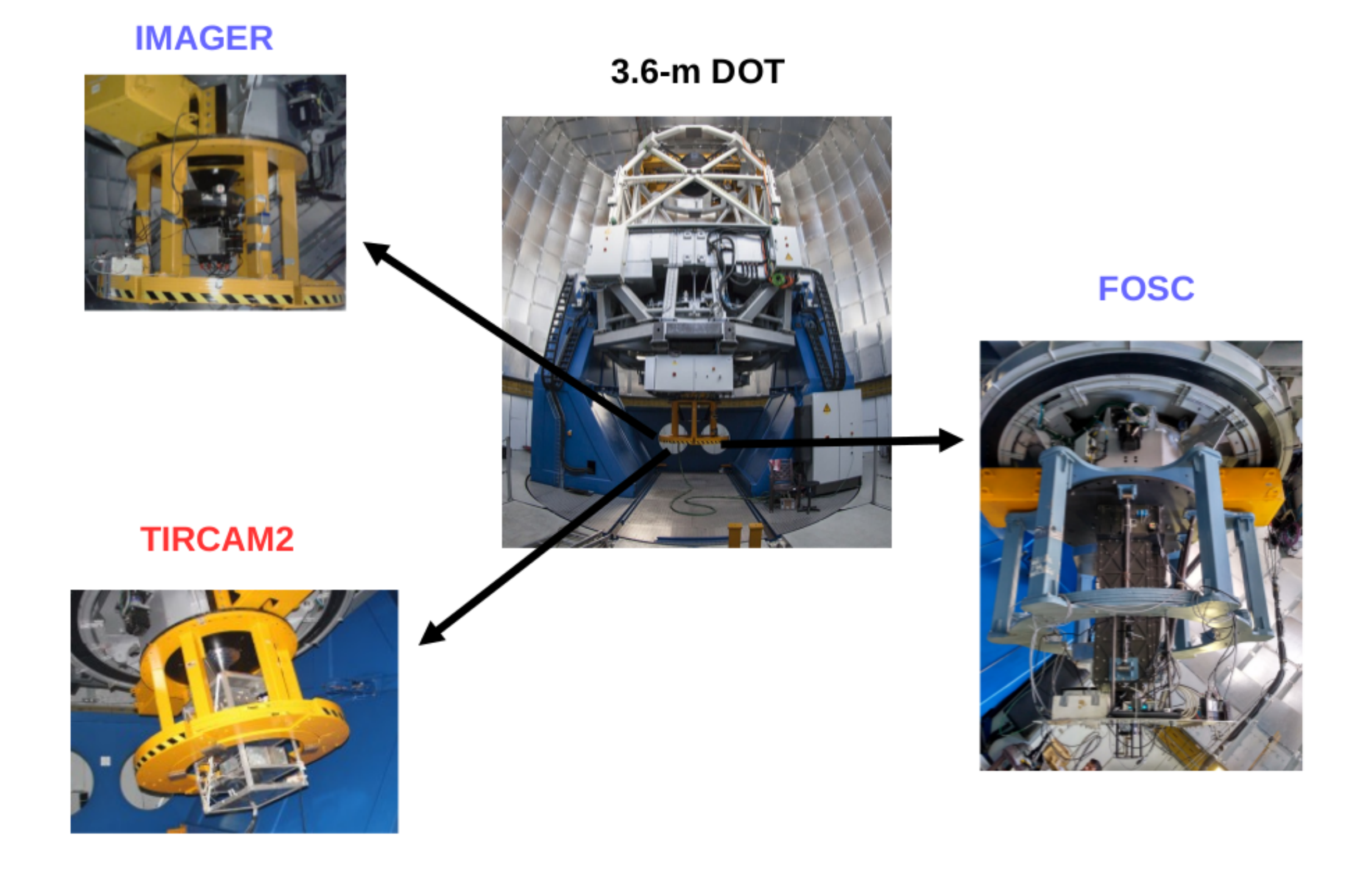}
\caption{ The as-built 3.6-m DOT is in the center. The pictures of back-end instruments 4K$\times$4K CCD IMAGER, 
Faint Object Spectrograph and Camera (FOSC) and TIRCAM2  used for regular observations are shown. \label{fig:fig-00a}}
\end{figure}

During commissioning, the performance of the 3.6-m DOT was tested using Test-Camera viz. an air-cooled Microline 
ML 402ME CCD chip of 768$\times$512 pixels having a pixel size of 9 \micron\  which corresponds to 0\farcsec06  at the
f/9 focal plane. The images of a few binary stars with known separation were observed on different nights using 
broadband (0.45 to 0.6 \micron) visual glass filter. Subsequently, professional instruments were mounted on the 3.6-m
DOT and tested. Figure 2 shows the pictures of the as-built telescope along with back-end instruments being used 
for regular observations. For observations at optical wavelengths, 4K $\times$ 4K CCD IMAGER (Pandey et al. 2018) and 
Faint Object Spectrograph and Camera (FOSC) (Omar et al. 2012, 2019a) are used while TIRCAM2, acronym for TIFR Near 
Infrared Imaging Camera-II, is used in NIR region. The 3.6-m DOT was used for detailed characterisation of these 
instruments as well as for some science programs, which included photometric and spectroscopic study of supernovae, 
optical transient events, AGNs and quasars, galactic star forming regions, and distant galaxies. Omar et al. (2017) 
and Sagar (2017) have highlighted scientific potential of the 3.6-m DOT. The key results of 
the sky quantification are briefly summarised in the ensuing sub-sections.

\subsection{Capabilities at optical wavelengths}

The CCD sky images taken with an instrument are blurred due to the contributions from optics in the instrument, Earth's 
atmosphere seeing and drift in telescope tracking etc. The Test-Camera was used to resolve binary star having angular separation of 
0\farcsec4 (Kumar et al. 2018; Omar et al. 2017; Sagar et al. 2019) indicating that the performance of the telescope 
system combined with that of the site and telescope building 
is up to the mark. It indicates that the care taken in the construction of the telescope buildings has paid rich dividend 
as their thermal mass is so low that it has not degraded the natural atmospheric seeing at Devasthal measured about 
two decades ago during 1997 to 1999 using differential image motion monitor (DIMM).

\begin{figure} 
\centering
\includegraphics[width=12cm]{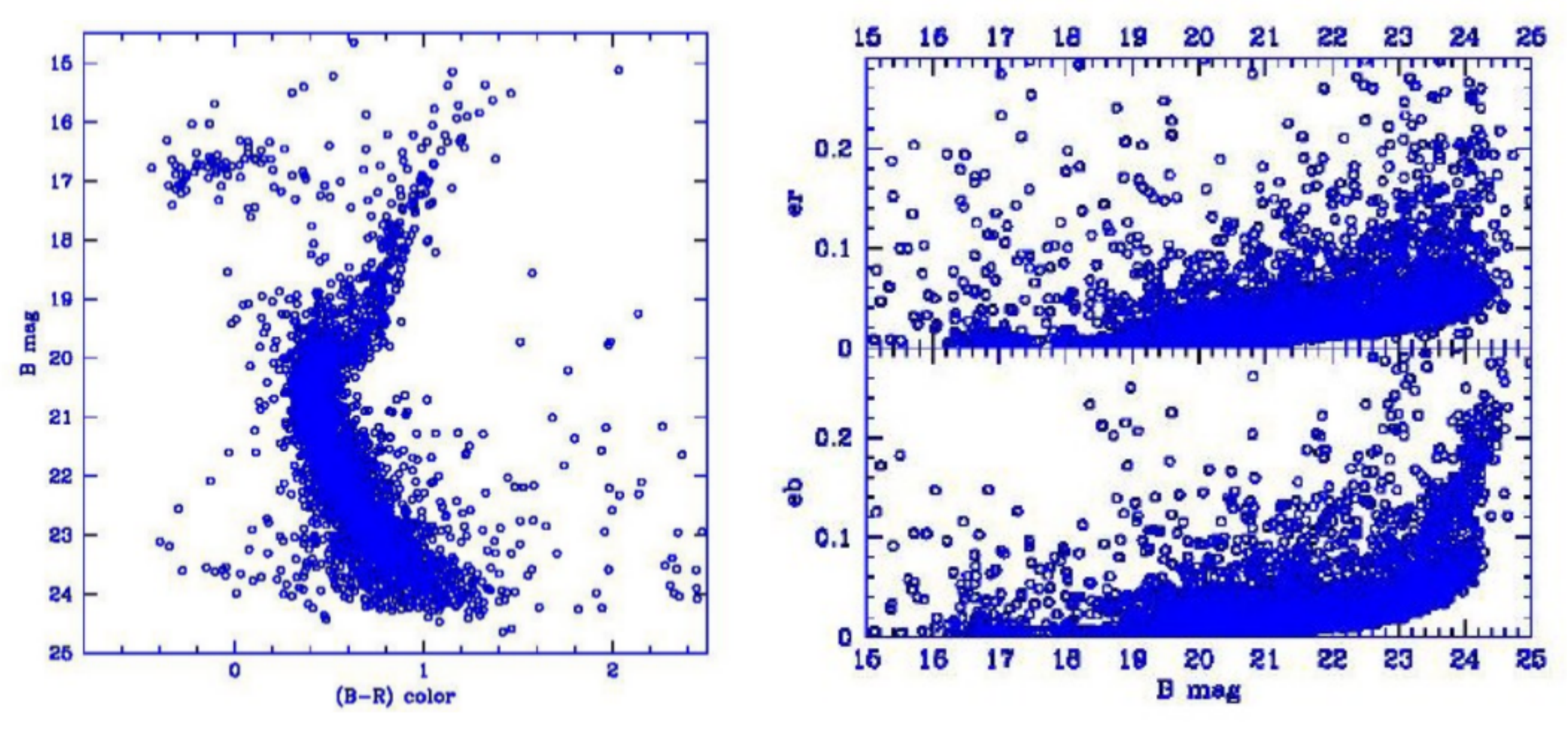}
\caption{Left panel displays the B versus (B-R) color-magnitude diagram (CMD) of the Galactic GC NGC 4147 obtained from the
images taken by Pandey et al. (2018) using 4K $\times$ 4K CCD IMAGER mounted on axial port of the 3.6-m DOT. Right panel
shows plot of photometric errors (mag) in both B (eb) and R (er) bands versus B magnitude. Most of the stars fainter than B= 24 mag
have photometric errors $\geq$ 0.2 mag.}
\end{figure}

Pixel size of a blue-enhanced liquid Nitrogen cooled STA4150 4K $\times$ 4 K CCD IMAGER is 15 \micron. It is
equipped with standard broadband UBVRI and SDSS $\it ugriz$ filters and covers $6\farcmin5 \times 6\farcmin5$ field of view 
when mounted on the 3.6-m DOT. Figure 3 shows a colour-magnitude diagram of the galactic globular cluster (GC) NGC 4147 
which is based on the images taken with the 4K$\times$4K CCD imager by Pandey et al. (2018). This indicates that stars fainter than 24 mag in B filter have been detected with an accuracy of $\sim$ 0.2 mag in an effective exposure time of 20 min. 
Using the FOSC, a 24.5 mag object has been detected with an accuracy of 0.2 mag in one hour of integration time in $i-$band 
(Omar et al. 2019b). The detection limits at optical wavelengths are at par with the performance of other 3.6 meter 
aperture optical telescopes located at good observing sites. The night sky brightness values are estimated to be 21 and 20.4 
(mag arcsec$^{-2}$) in the $r$ and $i$ bands respectively (Omar et al. 2019b). These numbers indicate that the night sky at 
Devasthal is dark and it has not degraded since late 1990’s when Devasthal site characterization was carried out. It 
is also at par with other major international optical observatory sites including Leh-Hanle (Prabhu 2014).

\subsection{Capabilities at NIR wavelengths}

The wavelength sensitivity of TIRCAM2, a closed cycle Helium cryo-cooled $512\times512$ pixels imaging camera, is
from 1 to 5 \micron. On the 3.6-m DOT, pixel scale of the camera  is 0\farcsec17. Ojha et al. (2018) and Baug et
al. (2018) provide more technical and performance details of this camera. It is equipped with standard $J (1.20 \micron)$, $H (1.65 \micron)$ and $K (2.19 \micron)$ broad $(\Delta \lambda \sim 0.3-0.4 \micron)$ photometric bands and with narrow 
 $(\Delta \lambda \sim 0.03-0.07 \micron)$ band $B_{r-\gamma} (2.16 \micron); K_{cont} (2.17 \micron)$; Polycyclic aromatic hydrocarbon  $(PAH$) ($3.28 \micron$) and narrow band L $(nbL) (3.59 \micron)$ filters. The camera is capable of observing sources up to 19.0, 18.8 and 18.0 mag with 10\%  photometric accuracy in $J$, $H$ and $K$ broad photometric bands respectively, 
 with corresponding effective exposure times of 550, 550 and 1000 sec respectively. It is also capable of detecting the $nbL$ band sources brighter than $\sim$ 9.2 mag and strong ($\geq$ 0.4 Jy) $PAH$ emitting sources like Sh 2-61. During good sky conditions, Baug et al. (2018) estimated 16.4, 14.0, 12.2 and 3.0 mag arcsec$^{-2}$ as sky brightness in $J, H, K$ and $nbL$ bands respectively which are comparable with other observatories like Hanle (Prabhu 2014), Calar Alto Observatory (Sanchez et al. 2008), South Pole (Ashley et al. 1996) and Las Campanas Observatory  (Sullivan \& Simcoe 2012).

\begin{figure}[h]
\centering
\includegraphics[width=16cm]{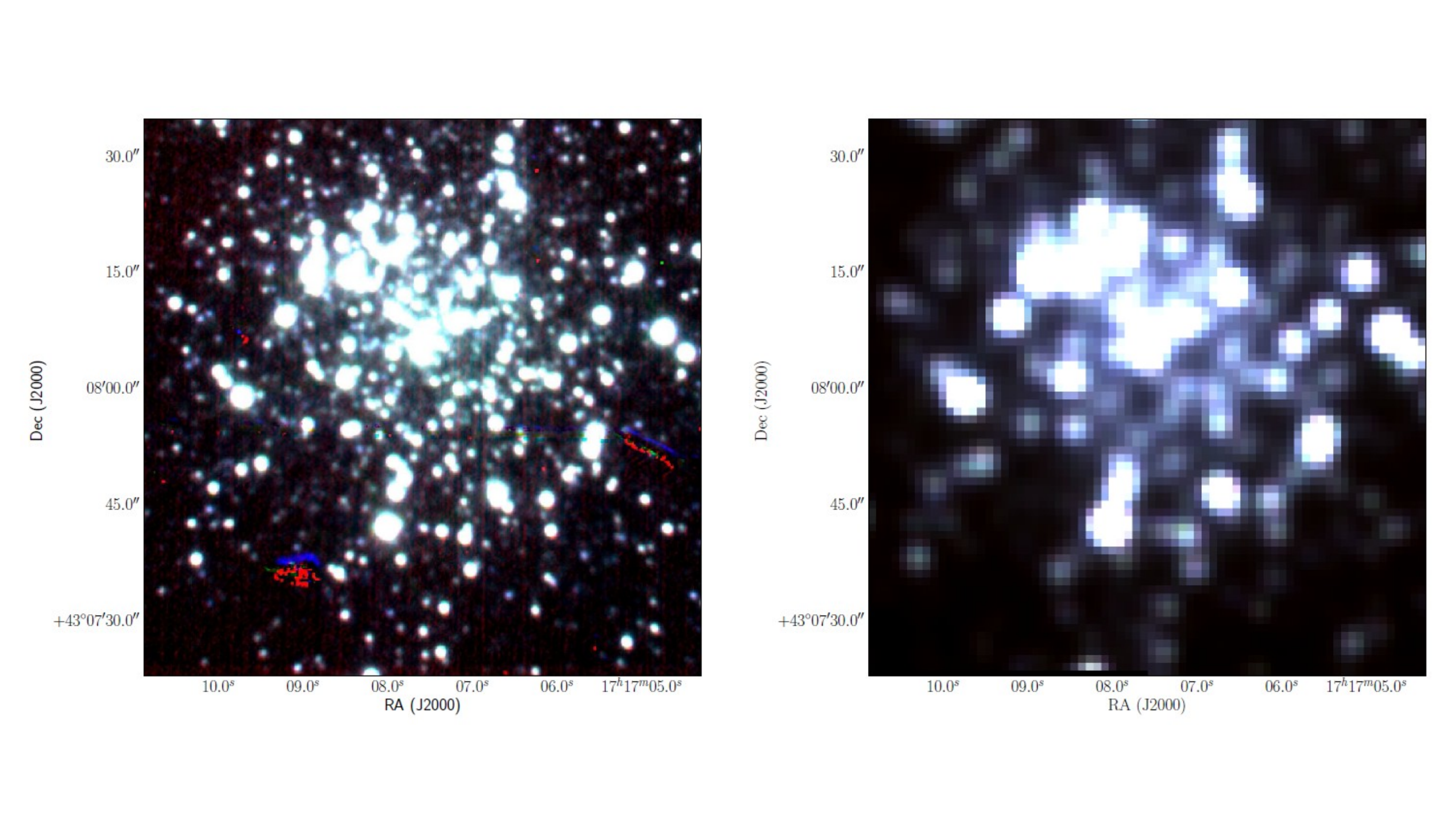}
\caption{Left panel shows the color-composite image of the  GC M 92 constructed from the $J$ (blue), $H$ (green) and $K$ (red) band frames observed with the TIRCAM2 mounted on the 3.6-m DOT. A color-composite image for the same area generated from the 2MASS $J, H,$ and $K$ band images is also shown in the right panel for comparison. Both images have field of view $\sim$ 100\arcsec$\times$70\arcsec and centred at $\alpha_{2000} = 17^h 17^m 7.5^s; \delta_{2000} = +43\arcdeg 08\arcmin$.}
\end{figure}

The Devasthal sky performance at NIR wavelengths has been characterized for the first time. Figure 4 shows a comparison 
of the $J, H$ and $K$ colour-composite images of the GC M 92 taken by Ojha et al. (2018) with the TIRCAM2 mounted on the 
3.6-m DOT with that generated from the the 2MASS $J, H$ and $K$ band images. This indicates that angular resolution of
NIR images obtained with 3.6-m DOT is better than that of 2MASS images. In the $K$ band, stellar images with $\sim$ 0\farcsec6 to 0\farcsec9 of full width at half maximum (FWHM) were obtained routinely while the best value of the FWHM ($\sim$ 0\farcsec45) was observed on 16 October 2017. The 3.6-m DOT is therefore adequate for deep NIR observations that are comparable to other 4 meter class telescopes available world-wide.

\section{Other Back-end Instruments on the 3.6-m DOT}

It is well known that modern and complex back end focal plane instruments are essential to get best scientific results from any telescope. Due to their great technological complexity, they are generally designed and developed using expertise available across institutions located anywhere in the globe. Development of two such instruments namely Devasthal Optical Telescope Integral Field Spectrograph (DOTIFS) and TIFR-ARIES Near Infrared Spectrometer (TANSPEC) for the 3.6-m DOT are in advanced stage. Present status of both instruments are presented in the following sub-sections.

\subsection { Devasthal Optical Telescope Integral Field Spectrograph (DOTIFS)}

DOTIFS is a multi-object Integral Field Spectrograph (IFS). It is being developed, designed and fabricated by the Inter-University Center for Astronomy and Astrophysics (IUCAA) as a part of an inter-institutional collaboration. Korea Institute for Advanced Study and Seoul National University are participating in this project as international collaborators. The DOTIFS consists of one magnifier,
16 integral field units (IFUs) and 8 high throughput spectrographs and will be mounted on the Cassegrain side port of 
the 3.6-m DOT. Technical details of the DOTIFS and its various components are published by Chung et al. (2014). In a single 
exposure, dedicated spectrographs will produce 2,304 spectra of $\sim$1800 resolution over a wavelength range of 
0.37 to 0.74 \micron. DOTIFS will use Volume Phase Holographic gratings for making optics smaller and also improving 
spectrograph throughput. The optical and opto-mechanical design of the DOTIFS spectrograph has been described by 
Chung et al. (2018a) while details of fore-optics and calibration units are given by Chung et al. (2018b). As per 
design, average throughput of the DOTIFS mounted on the 3.6-m DOT is expected to be over 25\%. 

\subsection {TIFR-ARIES Near Infrared Spectrometer (TANSPEC)}

The TANSPEC, an Optical-NIR medium resolution spectrograph, is jointly developed by TIFR and ARIES as a part of 
inter-institutional collaboration. The wavelength coverage of this unique spectrograph is from 0.55 $\micron$ in 
optical up to 2.54 $\micron$ in NIR with a resolving power of $\sim$2750 and a capability of simultaneous 
observations across entire wavelength region. Ojha et al.(2018) provide its present status along with 
technical details and optical lay out of this instrument. Briefly, it converts the f/9 telescope beam into 
f/12 beam on to the slit which has a range of slit widths from 0\farcsec5 to 4\farcsec0. Field of view of the slit 
viewer is 60 \arcsec  $\times 60 \arcsec $ while its one pixel corresponds to 0\farcsec25 on the sky.   
One pixel of the spectrograph 2048 X 2048 Hawaii-2RG (H2RG) array corresponds to 0\farcsec25 and it operates in 
two modes. In highest resolution ($\sim$ 2750) mode, combination of a grating and two prisms are 
 used while low resolution ($\sim$ 100) prism mode is used for high throughput observations. The instrument also 
 has an independent imaging camera with a $1 K \times 1 K$ H1RG detector which is the slit viewer.  This camera has a 
 field of view of $1 \arcmin \times 1 \arcmin$, and is used for the telescope guiding as well as for sky imaging. 
 It also functions as a pupil viewer for instrument alignment on the telescope. It is equipped with both broad
 $(r', i', Y , J, H, K_{s})$ and narrow $(H_{2} \& B_{r})$ band filters. The TANSPEC, after successful completion of laboratory test at TIFR, Mumbai, has been transported to Devasthal site and soon, it may become available for users as an important back-end instrument on the 3.6-m DOT. 

 \section {Key scientific areas in Galactic Astronomy}  

In the light of above, it can be said that observing facilities located at Devasthal coupled with its good sky 
conditions have potential of providing not only internationally competitive  but scientifically valuable optical 
and NIR observations for a large number of front line galactic astrophysical research problems. Geographical location 
of the Devasthal observatory also has global importance for the time domain and multi-wavelength astrophysical studies.

Since 2016, the 3.6-m DOT has been used for both optical and NIR observations of research proposals related to studies of
star formation, star clusters, stellar evolution, transient objects, variable stars and Exo-planets in the field of 
galactic astrophysics. Pandey et al. (2018) and Ojha et al. (2018) have provided a detailed description of these topics. 
TANSPEC and TIRCAM2 at the focal plane of the 3.6-m DOT will be a major workhorse for a variety of challenging 
astrophysical problems since they are sensitive to low temperature (T $\sim$ 2500 K) stellar atmospheres and objects 
surrounded by warm dust envelopes or embedded in dust/molecular clouds. Hence, they are very much suited for the 
search of low and very low mass stellar populations (M dwarfs, brown dwarfs), strong mass-losing stars on the asymptotic 
giant branch and young stellar objects still in their proto-stellar envelopes.

We have taken a research project entitled study of Ultra-Violet (UV) stars in star clusters of our galaxy. For this, we have obtained UV observations from the India$^{'}$s space Observatory, {\bf AstroSat} and are using the 3.6-m DOT for optical and NIR observations. They are briefly presented in the following sub-sections along with capabilities of UV imaging telescope (UVIT) payload.

 \subsection{ The UVIT payload on the AstroSat Satellite} \label{astrosat}
 
 \begin{figure*} [h]
\centering
\includegraphics[scale=0.5]{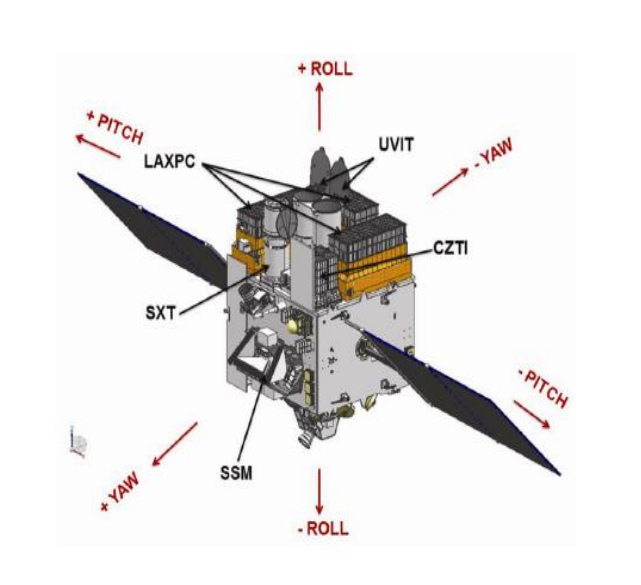}
\caption{ This figure shows locations of all five payloads on the space craft. It is taken from the ISRO public website   http://www.isro.gov.in/astrosat/astrosat-gallery. The payloads are Ultraviolet Imaging Telescope (UVIT), Soft X-ray Telescope (SXT), Scanning Sky Monitor (SSM), Large Area X-ray Proportional Counter (LAXPC) and Cadmium-Zink-Telluride Imager (CZTI). Definition of 
the three space craft axes (ROLL, YAW and PITCH) are elaborated by Seetha \& Megala (2017). \label{fig_3}}
\end{figure*}
 
 {\bf AstroSat}, India$^{'}$s first multi-wavelength space-born astronomy mission launched successfully on 
 28 September 2015, is capable of carrying out simultaneous multi-wavelength observations of celestial bodies 
 ranging from X-ray to optical since it caries 5 scientific payloads (Figure 5). One of them is the UV 
 Imaging Telescope (UVIT). Seetha \& Megala (2017) have provided more detailed information on this space observatory 
 while Tandon et al. (2017) have presented the capabilities and on-board performance of the UVIT payload. The UVIT is 
 a combination of two co-aligned f/12 Ritchey-Chretien reflecting telescopes of aperture 375 mm. One of them is 
 equipped with a Far Ultraviolet (FUV) detector while other is built to observe Visible (VIS) and Near Ultraviolet 
 (NUV) part of the electromagnetic radiation by using a dichroic beam splitter which splits NUV and VIS. The modern 
 detectors mounted at the telescope focii provide a wide circular field of view $\sim$ 28\arcmin\ diameter and high 
 angular resolution of $\sim$ 1\farcsec2. The number of filters in FUV, NUV and VIS are 5, 6 and 5 respectively.  
 A few scientific results based on the UVIT observations taken by our group are mentioned below.
 
 \subsubsection{ Far Ultraviolet UVIT images of Galactic globular clusters NGC 1851 and NGC 288}
 
\begin{figure*} 
\centering
\includegraphics[scale=0.5]{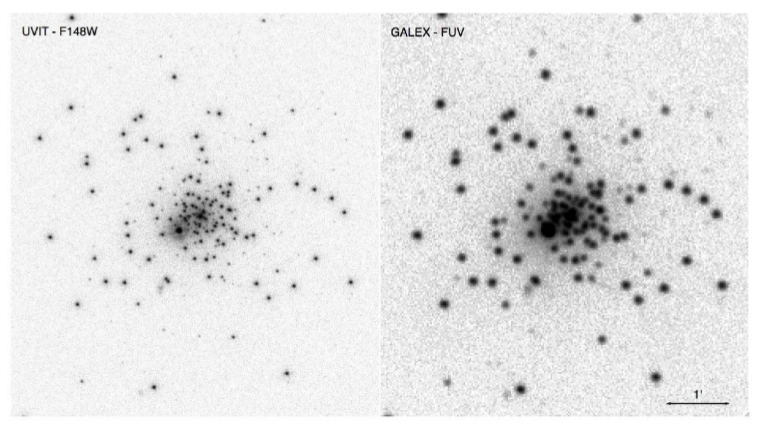}
\caption{Comparison of a globular cluster NGC 1851 UVIT FUV (F148W filter) image of 6982.13 sec integration with that of a GALEX FUV image. North is up and east is to the left. Both images are centered at $\alpha_{2000} = 5^h 14^m 6.^{s}76; \delta_{2000} = - 40\arcdeg 02\arcmin 47\farcsec6$ and have field of view $6\arcmin \times 6\farcmin5 $. It shows that angular resolution ($\sim 1\farcsec2$) of UVIT image is $\sim$ 4 times better than that ($\sim 5\arcsec$) of GALEX image. \label{fig_4}}
\end{figure*}

  In Figure 6, the UVIT FUV image of a massive southern GC NGC 1851, taken by  Subramaniam et al. (2017), is compared with that of an image taken with GALEX space mission. Angular resolution of the UVIT image is $\sim$ 1\farcsec2 which is $\sim$ 4 times better in comparison to GALEX images as it has resolution of $\sim$ 5\arcsec. Consequently, stars located in the core of the cluster are almost  resolved in the UVIT NUV and FUV images and their precise magnitudes could be determined using point spread function techniques of stellar photometry in crowded regions. The good image quality combined with large field of  view make the UVIT extremely valuable for studies of hot stellar populations in GCs  since it is able to detect the UV and blue stars beyond the central region of the cluster.  Based on FUV and NUV images taken with the UVIT, Sahu et al. (2019) have presented a census of the blue horizontal branch (BHB) and Blue Straggler stars (BSSs) population in the GC NGC 288 while Subramaniam et al. (2017) have studied it for the NGC 1851. These scientific results prove beyond doubt that the UVIT images, in combination with the Hubble Space Telescope (HST), ground and GAIA DR2 data,  are a very powerful tool to understand the formation of UV and blue stellar population in the core of galactic GCs. Combination of the superior image quality and the number of filter systems in the both FUV and NUV regions of the UVIT have provided valuable information about the spatial distributions of BHB and BSSs in the both GCs NGC 1851 and NGC 288.  

 \subsubsection{ Discovery of a rare type of Blue straggler star in NGC 188 }

\begin{figure*} 
\centering
\includegraphics[scale=0.65]{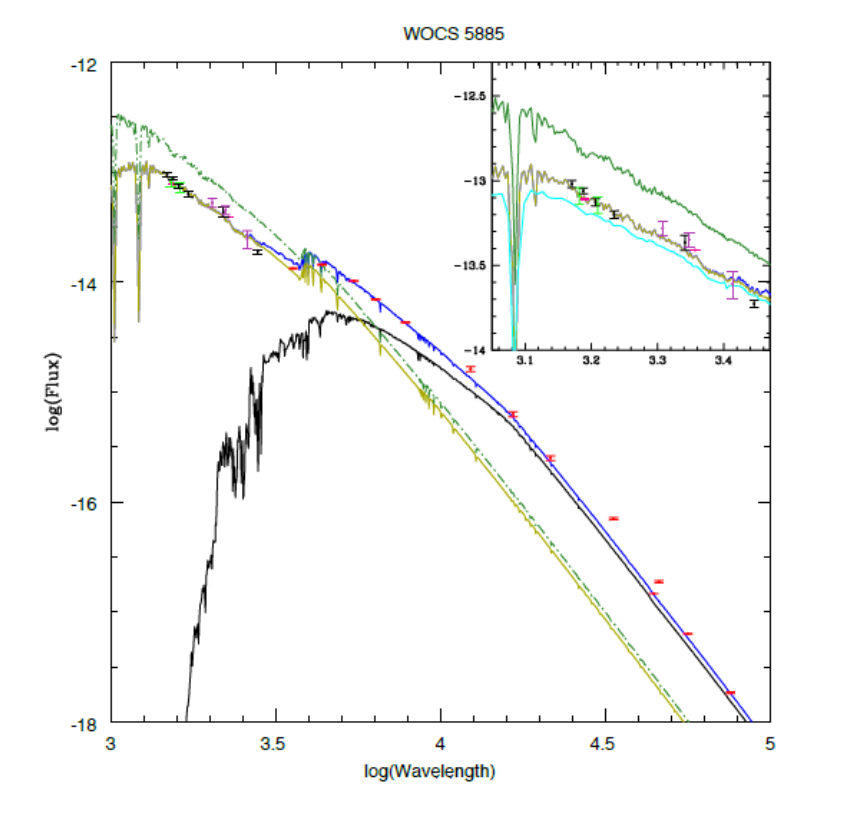}
\caption{The extinction corrected spectral energy distribution of the blue straggler WOCS 5885 is borrowed
from Subramaniam et al. (2016). The black (UVIT), magenta (GALEX) and green (UIT), pink (UVOT) points 
indicate the UV fluxes (shown in the inset as well); all other flux measurements are shown in red. Kurucz 
Model spectra (Log(g) = 5.0) for the separate components are shown in gold (17,000 K) and black (6,000 K), 
with the composite spectrum in blue. Scaling factors of 4.45E-22 and 3.1E-23 have been used to combine 
the 6,000 K and 17,000 K spectra,respectively. The unit of wavelength is Angstrom and flux is 
ergs $cm^{-2} s^{-1} A^{-1}$. \label{fig_5}}
\end{figure*}
 
The BSS WOCS 5885 in the old ($\sim$ 7 Gyr) open star cluster NGC 188, is found to be one of the brightest 
FUV sources in the cluster. The spectral energy distribution (SED) of the star WOCS 5885 has been constructed 
by Subramaniam et al. (2016) by combining fluxes measured in four FUV and two NUV UVIT filters with flux 
measurements from other space and ground-based observatories (Figure 7). The resulting SED spans a wavelength 
range of 0.15 to 7.8 $\mu$m. A single temperature was unable to fit the full SED. The UV part of the SED seems 
to suggest a relatively hot temperature; whereas the NIR part suggests a cooler temperature.
The values of two temperatures are estimated to be 17,000($\pm$500) K and 6,000($\pm$150) K 
for the hotter and cooler components respectively. The optical region of the SED is fitted well by 
the combined flux from the two temperature components. The high temperature component is revealed only by 
the UVIT flux measured using the F148W and F154W filters (inset in Figure 7 suggests that the UV flux is 
rising at least until 1481 \AA). The slope of the FUV part of the SED indicates that the temperature of the hot 
component is not high enough to be classified as a sub-dwarf. The estimated temperature of the cooler 
component (BSS) is similar to the temperatures of other BSSs in NGC 188. Therefore, Subramaniam et al. (2016) 
proposed that the hot component (17,000$\pm$500 K) is a post-AGB/HB star that has recently transferred its 
mass to the BSS. The WOCS 5885, the first BSS with a post-AGB/HB companion identified in an open star cluster, 
can be considered an ideal object to study the process of BSS formation via mass transfer in a binary system.

\section{Summary and Conclusions}

On sky performance of the 3.6-m DOT reveals that quality of its optics is excellent and capable of providing 
images of the celestial bodies with sub-arc-second (up to 0\farcsec4) resolutions. CCD images of a star cluster 
taken with the 3.6-m DOT indicate that stars fainter than 24$\pm \sim$0.2 mag in $B$ are detected during 20 minutes of exposure time while in $i-$band a star of 24.5 mag has been detected with a photometric precision of 0.2 mag in one hour of integration time. The detection limits at optical wavelengths are thus at par with the performance of other similar optical telescopes located at good observing sites. 

An average seeing (FWHM) of $\sim 0\farcsec7$ observed in $K$ band correspond to $\sim 0\farcsec9 $ at visual bands since atmospheric seeing varies as $ \lambda^{-0.2}$. Observing sub-arc-second atmospheric seeing at Devasthal even after completion of the telescope and surrounding infrastructure buildings  indicates that they have not deteriorated natural seeing conditions observed at the site using DIMM about two decades ago during 1997 to 1999. It is mainly because the precautions taken in the design and the structure of the telescope house have kept their thermal mass very low. Both optical and NIR observations indicate that 3.6-m DOT can provide sky images with sub-arc-sec resolution for a good fraction of observing time. The sky performance of Devasthal at NIR wavelength, known only recently with the TIRCAM2 observations, is very encouraging. 

As discussed above, the telescope collects photons while the throughput of back end instruments defines the quality of scientific output coming from the telescope. The 4 meter class modern observing facilities located at Devasthal have advantage of their geographical location for providing optical and NIR observations of not only transient objects like GRB afterglows and Supernovae but also of astrophysically significant $\gamma$-ray, X-ray, UV and radio sources observed with Indian and other observatories operating at these wavelengths. In the field of galactic astrophysics, the 3.6-m DOT is being used for both optical and NIR observations of research proposals related to studies of star formation, star clusters, stellar evolution, transient objects, variable stars and Exo-planets. TANSPEC and TIRCAM2 at the focal plane of the 3.6-m DOT are very much suited for the search of low and very low mass stellar populations (M dwarfs, brown dwarfs), strong mass-losing stars on the asymptotic giant branch and young stellar objects still in their proto-stellar envelopes.

 The UVIT payload on {\bf AstroSat} has global importance as FUV UVIT images have good angular resolutions ($\sim$ 1\farcsec2) over a wide circular region ($\sim$ 28\arcmin\ diameter) of the sky. Also, it has number of FUV and NUV filters in comparison to other recent similar space missions. UVIT images, in combination with other space and ground based data, are a very powerful tool to understand the formation of UV and blue stellar population in star clusters (see Sahu et al. 2019 and Subramaniam et al. 2016, 2017). However, for proper understanding of their progenitors, extensive photometric and spectroscopic studies at optical and NIR wavelengths are essential. It is therefore one of the key science project for the 3.6-m DOT in the field of galactic astronomy and astrophysics. Our group is actively involved in this project and was allotted observing time on the 3.6-m DOT.

%
\section*{Acknowledgements}
 Authors are thankful to the Reviewer for providing constructive comments which improved presentation of the manuscript. Help provided by the staff of both the IIA UVIT payload operations center and the Devasthal Observatory, Nainital is thankfully acknowledged. RS thanks the National Academy of Sciences, India (NASI) for award of Fellowship and the Director, IIA, Bengaluru for providing necessary infrastructural support. He also acknowledges the local support provided by the Scientific Organizing Committee of the 2nd BINA workshop and the travel support given by the Alexander von Humboldt Foundation, Germany through its group linkage research program.
     
%
%
%

\footnotesize
\beginrefer

\refer Ashley M. C. B., Burton M. G., Storey J. W. V. et al. 1996, PASP, 108, 721

\refer Baug T., Ojha D.K., Ghosh S.K. et al. 2018, JAI, 7, 1850003 DOI: 10.1142/S2251171718500034

\refer Bheemireddy K., Gopinathan M., Pant, J. et al. 2016, Proceedings of the SPIE conference,  9906, E44B, id. 990644 DOI:10.1117/12.2234727

\refer Chung H., Ramaprakash A. N., Omar A. et al. 2014, Proceedings of the SPIE conference, 9147, 
        id 91470V  DOI: 10.1117/12.2053051
 
 \refer Chung H., Ramaprakash A. N., Khodade P. et al. 2018a, Proceedings of the SPIE conference, 10702,  id 107027A DOI: 10.1117/12.2311594
 
 \refer Chung H., Ramaprakash A. N., Khodade P. et al. 2018b, Proceedings of the SPIE conference, 10702, id 107027U DOI: 10.1117/12.2312394
 
\refer De Cate P., Surdej J., Omar A. et al. 2018,   BSRSL, 87, 1

 \refer Flebus C., Gabriel E., Lambotte S. et al. 2008, Proceedings of the SPIE conference, 7012, 
           id. 70120A DOI: 10.1117/12.787888

\refer Kumar B., Omar A., Gopinathan M. et al. 2018, BSRSL, 87, 29

\refer Ninane N., Flebus C., Kumar B. 2012, Proceedings of the SPIE conference, 8444, 
              id. 84441V DOI:10.1117/12.925921

\refer Ojha D.K., Ghosh S.K., Sharma S. et al. 2018, BSRSL, 87, 58

\refer Omar A., Yadav R.K.S., Shukla V., Mondal S., Pant, J. 2012, Proceedings of the SPIE conference,  8446,  id. 844614 DOI:10.1117/12.925841

\refer Omar A., Kumar B., Gopinathan M., Sagar R. 2017, CSci, 113, 682

\refer Omar A., Kumar T.S., Krishnareddy B., Pant, J., Mahto, M. 2019a, CSci, (in press)/astroph-2019arXiv190205857O

\refer Omar A., Saxena A.,  Chand, K. et al. 2019b, JApA, 40, Art. 9 (6 p) https://doi.org/10.1007/s12036-019-9583-4

\refer Pandey S. B., Yadav R. K. S., Nanjappa, N. et al. 2018, BSRSL, 87, 42

\refer Pant P., Stalin C. S., Sagar R. 1999, A\&AS, 136, 19

\refer Prabhu T. P. 2014, PINSA, 80, 887 doi:10.16943/ptinsa/2014/v80i4/55174 

\refer Sagar R. 2017, PNAS, 87, 1 doi:10.1007/s40010-016-0287-8  

\refer Sagar R. 2018, BSRSL, 87, 391

\refer Sagar R., Stalin C. S., Pandey A. K. et al. 2000a, A\&AS, 144, 349

\refer Sagar R., Uddin W., Pandey A. K. et al. 2000b, BASI, 28, 429

\refer Sagar R., Omar A., Kumar B. et al. 2011, CSci, 101, 1020

\refer Sagar R., Kumar B. and Omar A. 2019, CSci, (under review)

\refer Sahu S., Subramaniam A., Cote P., Rao N.K., Stetson, P. 2019, MNRAS, 482, 1080

\refer Sanchez S. F., Thiele U., Aceituno J. et al. 2008, PASP, 120, 1244

\refer Seetha S., Megala S. 2017,  CSci, 113, 579

\refer Stalin C. S., Sagar R., Pant P. et al. 2001, BASI, 29, 39

\refer Semenov A. P. 2012,Proceedings of the SPIE conference, 8450,  id. 84504R DOI:10.1117/12.924645

\refer Subramaniam A., Sindhu N., Tandon S. N. et al. 2016, ApJ, 883, Article id. L27
    
\refer Subramaniam A., Sahu S., Postma J. E. et al. 2017, AJ, 154, Article id. 253

\refer Sullivan P. W., Simcoe R. A. 2012, PASP, 124, 1336

\refer Surdej J., Hickson P., Borra H. et al. 2018, BSRSL, 87, 68

\refer Tandon S. N., Hutchings J. B., Ghosh S. K. et al. 2017, JApA, 38, Article id. 28

\endrefer           

\end{document}